\documentclass[aps,superscriptaddress,twocolumn]{revtex4}
\usepackage{amsmath,amssymb,graphics,graphicx,dcolumn,bm,enumerate}
\usepackage{mathbbol}

\newcommand{\tcmp}
{\affiliation{Theoretical Condensed Matter Physics Division,
Saha Institute of Nuclear Physics,\\ 1/AF Bidhannagar, Kolkata 700 064 India.}}

\newcommand{\camcs}
{\affiliation{Centre for Applied Mathematics and Computational Science,
Saha Institute of Nuclear Physics, 1/AF Bidhannagar, Kolkata 700 064 India.}}

\newcommand{\isi}
{\affiliation{Economic Research Unit, Indian Statistical Institute, 203 B. T. Road, Kolkata 700 018, India.}}

\newcommand{\theory}
{\affiliation{Theory Division,
Saha Institute of Nuclear Physics, 1/AF Bidhannagar, Kolkata 700 064 India.}}

\newcommand{\cpt}
{\affiliation {Centre de Physique Th\'{e}orique (CNRS UMR 6207), Universit\'{e} de la M\'{e}diterran\'{e}e Aix Marseille II,
Luminy, F-13288 Marseille cedex 9, France.}}

\begin{document}

\title{Continuous transition of social efficiencies in the stochastic strategy minority game}

\author{Soumyajyoti Biswas}
\email[Email: ]{soumyajyoti.biswas@saha.ac.in}

\author{Asim Ghosh}
\email[Email: ]{asim.ghosh@saha.ac.in}
\tcmp

\author{Arnab Chatterjee}
\email[Email: ]{arnab.chatterjee@cpt.univ-mrs.fr}
\cpt

\author{Tapan Naskar}
\email[Email: ]{tapan.naskar@saha.ac.in}
\theory

\author{Bikas K. Chakrabarti}
\email[Email: ]{bikask.chakrabarti@saha.ac.in}
\tcmp 
\camcs 
\isi


\begin{abstract}
We show that in a variant of the minority game problem, the agents can reach a state of maximum
social efficiency, where the fluctuation between the two choices is minimum, by following a simple stochastic strategy.
By imagining a social scenario where the agents can only guess about the number of excess people in the majority, we
show that as long as the guessed value is sufficiently close to the reality, the system can reach a state of full
efficiency or minimum fluctuation. A continuous transition to less efficient condition 
is observed when the guessed value becomes worse. Hence, people can optimize their guess for excess 
population to optimize the period of being in the majority state. We also consider
the situation where a finite fraction of agents always decide completely randomly (random trader) as opposed to 
the rest of the population who follow a certain strategy (chartist). For a single random trader the system becomes fully efficient 
with majority-minority crossover occurring every 2 days on average. For just two random traders, all the agents
have equal gain with arbitrarily small fluctuations.
\end{abstract}

\maketitle

\section{Introduction}\label{sec:1}
The minority game (MG)~\cite{Challet:1997,Challet:1999,MGBook,Moro:2004} is a game of repeated choice of $N$ (odd) non-communicating 
agents where the agents ending up in the minority
receive positive pay-offs. This is a variant of the El Farol bar problem~\cite{ba}.   
Existence of a phase transition has already been reported in the MG~\cite{Challet:1999}.

One of the aims of the study of MG is to bring the populations in either choices sufficiently close 
to $(N-1)/2$ and in doing so to keep the time of convergence and fluctuations low~\cite{MGBook,Moro:2004}. As is
readily seen, a random choice would make the time of convergence virtually zero, but the fluctuation would be
of $\sqrt{N}$ order. This makes the system socially very inefficient. This is because, although $N/2$ agents could be 
in the minority at the same time, due to fluctuation this number may be much smaller than that, thereby making
the system socially inefficient due to poor resource utilization. To increase social efficiency, this fluctuation has to be minimised. 
Several adaptive strategies were studied to improve upon this situation. However,
regarding the fluctuation, even for the most complex strategies only the pre-factor could be made smaller 
and the fluctuation still scales with $\sqrt{N}$. A large misuse of resource is therefore likely in this situation.

There have been some recent attempts to study a generalization of the MG problem in the Kolkata paise restaurant (KPR)
problem, where there are $N$ agents and $N^\prime$ choices (restaurants). Keeping a finite comfort level (equal in most cases)
for the restaurants (say, one agent per choice) one arrives at similar problem of finding the most efficient strategy (non-dictated) 
for resource allocation~\cite{Chakrabarti:2009,demo,Ghosh:2010}. 
It was shown that a simple crowd-avoiding, stochastic, non-dictated strategy led to the most
efficient resource allocation for the problem in a very short convergence time (practically independent 
of the system size)~\cite{Chakrabarti:2009,Ghosh:2011}.

Stochastic strategies have been used also for the MG problem \cite{Reents:2001}. Recently, Dhar {\it et al.}~\cite{Dhar:2011} 
used the KPR strategy in the MG problem
to find that the fluctuation could be made arbitrarily small in a
 convergence time of the order $\log\log N$. 
This, of course, is the best possible strategy so far in terms of resource allocation in the MG. 
However, this situation differs a bit from the classic MG problem. The most important difference is that, being
a crowd-avoiding strategy, it requires that the agents know not only whether they were in the minority or majority
the previous evening, but also the number of excess people in the majority if he or she were there.

The present study  intends to determine if knowing the number of excess  is indeed necessary in reaching a state with minimum
fluctuations. We show that even in a more realistic situation where  exact knowledge of the number of excess people
is not available to the agents, it is possible to reach a  state with minimum fluctuations. A natural relaxation of 
the complete knowledge of the excess crowd would be to make a guess about the actual value. We show analytically 
for an idealized situation, where all agents make the same guess, that if the guess value is smaller than twice the real value,
the minimum fluctuation state (or the absorbing state) can be reached. However, for any larger error, a residual fluctuation
proportional to the excess error made, stays in the system in the steady state. Further, in a generalized and more
realistic version, where the guesses differ and are random for each agent in each step (annealed disorder), we 
analytically and numerically find similar
behavior in the fluctuations in the system only in terms of the average guess value this time. We show that 
depending upon the guessing power of the agents, there is
a continuous phase transition between full (minimum fluctuation or absorbing phase) and partial 
(system with residual fluctuation or active
phase) efficiencies for resource allocation in this problem. We also analyze the case where the knowledge of the excess population is completely unknown to the agents. It is shown both analytically and numerically that zero fluctuation can be reached by following an annealing schedule. To further incorporate the effects of real situations,
we consider the fraction of  agents who decide completely randomly. When the number of random traders is 1, although the fluctuation can be at its minimum, he or she is always in the majority. However, for more than one random traders, this situation can be avoided, maintaining a minimal fluctuation.

The rest of the paper is organized as follows: In Sec.~\ref{sec:2}, we discuss different strategies followed by the agents to
minimize the fluctuation. We show that by tuning a suitable parameter, an active-absorbing phase transition takes place. We elaborate on
other strategies where the agent's knowledge of the excess population is only partial or is completely absent. In Sec. III we discuss the effects of random traders.
Finally, in Sec. IV, we give concluding remarks.

\section{Strategy of the agents}\label{sec:2}
In the strategy followed in Ref.~\cite{Dhar:2011}, the agents in the majority shift with the probability
\begin{equation}
p_+=\frac{\Delta(t)}{M+\Delta(t)+1},
\label{strategy}
\end{equation}
and the agents in the minority remain with their choice ($p_-=0$), where the total population ($N=2M+1$) is divided as $M+\Delta(t)+1$ and $M-\Delta(t)$
with $\Delta(t)=(|N_A(t)-N_B(t)|-1)/2$, where $N_A(t)$ and $N_B(t)$ are the populations in the two choices at time $t$. 
In this strategy, the agents can reach the zero fluctuation limit in $\log \log N$ time \cite{Dhar:2011}. Although the
resource utiliszation is maximum in that case, its distribution is highly asymmetric in the sense that after the dynamics 
stops in the $\Delta(t)=0$ limit, the agents in the minority (majority) stay there for ever; hence, only one group  always benefits.  
Apart from this, in this strategy 
the knowledge of $\Delta(t)$ is made available to all the agents,
which is not a general practice in MG. 

In the following subsections, we introduce several variants of the above mentioned strategy. Primarily we intend to
find if it is possible to avoid the freezing of the dynamics while keeping the fluctuation as low as possible. We then discuss 
if it is possible to achieve
such states without knowing the magnitude of $\Delta(t)$. 
\subsection{Uniform approximation in guessing the excess crowd}
Let us assume that the agents know the value of $\Delta(t)$. We  then intend to find a strategy where
the dynamics of the game does not stop and the fluctuation can be made as small as required. 

To do that we propose the following strategy:
The shifting probability of the agents in majority is
\begin{equation}
p_+(t)=\frac{\Delta^{\prime}(t)}{M+\Delta^{\prime}(t)+1},
\end{equation}
[where $\Delta^{\prime}(t)=g\Delta(t)$ and $g$ is a constant] and that from the minority remains zero. We will see that a steady state can be reached in this model where the fluctuation can be
arbitrarily small.

\subsubsection{Steady-state behavior}
Now, to understand when such a steady state value is possible, note that when the transfer of the crowd from
majority to minority is twice the difference of the crowd, the minority then will become the majority and will have
the same amount of  excess people as  in the initial state. Quantitatively, if the initial populations
were $M+\Delta$ and $M-\Delta$ roughly, and if $2\Delta$ is shifted from majority to minority, then the situation
would continue to repeat itself, as the transfer probability solely depends on the excess crowd. Of course, this is 
possible only when $g>1$. More formally,
if the steady-state value of $\Delta(t)$ is $\Delta_s$, then the steady state condition would require
\begin{equation}
(M+\Delta_s+1)\frac{g\Delta_s}{M+g\Delta_s+1}=2\Delta_s.
\end{equation}
This, on simplification yields either $\Delta_s=0$ or
\begin{equation}
\Delta_s=\frac{g-2}{g}(M+1).
\end{equation}
\begin{figure}[tb]
\centering \includegraphics[width=8.5cm]{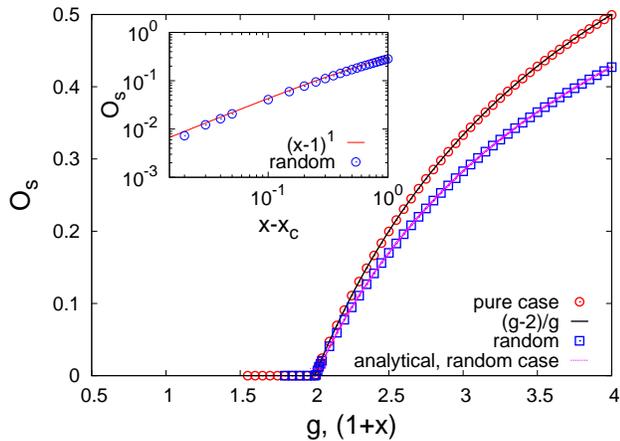}
   \caption{Steady state values of the order parameter $O_s$ for different values of $g$ and $x$. The solid lines shows the analytical results for the pure and annealed disordered cases. Both match very well with the simulation points.  Inset shows the log-log plot near the critical point for the disordered case, confirming $\beta=1.00\pm0.01$.
All  simulation data are shown for $M=10^5$.}
\label{op}
\end{figure}
Clearly, for $g<2 (=g_c)$, $\Delta_s=0$ would be the valid solution, since the above equation predicts a negative value for $\Delta_s$,
which indicated no steady-state saturation until it decreases to zero. Therefore, one can predict a phase transition
of the active-absorbing type \cite{lubek} by tuning the value of $g$. When $0<g<2$, the system will reach the minimum fluctuation state
where $\Delta(t)=0$ and the dynamics stops (the dynamics will differ qualitatively  for $g<1$ and $g>1$; see the appendix). 
For $g>2$, however, a residual fluctuation will remain in the system, keeping it active.
Physically, this would mean that until the guessed value of the crowd is not too incorrect (twice as large), the agents can still find
the minimum fluctuation state. However, when the guess becomes too wild, a fluctuation remains in the system. 

Therefore, it is now possible to define an order parameter for the problem as $O(t)=\Delta(t)/M$ and its saturation values
behaves as $O_s=0$ when $g<2$ and when $M\gg 1$, for $g>2$, with $O_s=(g-g_c)/g$ giving the order parameter exponent $\beta=1$ for this continuous transition. 
In Fig.~{\ref{op}}  we plot the results of numerical simulation ($M= 10^5$) as well as the analytical expression for the order parameter.
We find satisfactory agreement. 

\subsubsection{Dynamics of the system}
In this simple situation, it is possible to calculate the time dependent behavior of the order parameter both
at and above the critical point. Suppose, at a given instant $t$, the populations in the two choices $A$ and $B$ are
$N_A(t)$ and $N_B(t)$ respectively with $N_A(t)>N_B(t)$. By definition 
\begin{equation}
\Delta(t)=\frac{N_A(t)-N_B(t)-1}{2}.
\end{equation}
Moreover, the amount of the population to be shifted from $A$ to $B$ using this strategy would be
\begin{eqnarray}
\label{trns}
S(t) &=& \frac{g\Delta(t)}{M+g\Delta(t)+1}(M+\Delta(t)+1) \nonumber \\
&\approx & g \Delta (t), 
\end{eqnarray}
when $\Delta(t)$ is small compared to $M$, i.e., when $g$ is close to $g_c$ or for large time if $g\le g_c$. 

Clearly, $N_A(t+1)=N_A(t)-S(t)$ and $N_B(t+1)=N_B(t)+S(t)$, giving (assuming population inversion; see the appendix for a general treatment)
\begin{eqnarray}
\Delta(t+1) &=& \frac{N_B(t+1)-N_A(t+1)-1}{2} \nonumber \\
&\approx & g\Delta(t)-\Delta(t)-1.
\end{eqnarray}
Therefore, the time evolution of the order parameter reads
\begin{equation}
\frac{dO(t)}{dt}=-(2-g)O(t)-\frac{1}{M}.
\label{diffO}
\end{equation}
Neglecting the last term and integrating, one gets
\begin{equation}
O(t)=O(0)\exp[-(2-g)t].
\end{equation}
The above equation signifies an exponential decay of the order parameter in the subcritical region ($1<g<2$).
It also specifies a time scale as $\tau \sim (g_c-g)^{-1}$ diverging as the critical point is approached. These behaviors are confirmed by
numerical simulations. 

In Eq.~(\ref{trns}), the approximation was made to keep the leading order term only. If, however, the first correction
term is  kept, the expression becomes
\begin{equation}
S(t) \approx g\Delta(t)-\frac{1}{M}(g^2\Delta^2(t)-g\Delta^2(t)).
\end{equation}
The time evolution equation of the order parameter then becomes
\begin{equation}
\frac{dO(t)}{dt}=-(2-g)O(t)-g(g-1)O^2(t)-\frac{1}{M}.
\label{diff1}
\end{equation}
Now, if we consider the dynamics exactly at the critical point, i.e., $g=2$, then the first term in the right-hand-side is zero. The last
term can be neglected. Therefore, the order parameter becomes
\begin{equation}
O(t)=\frac{O(0)}{2 O(0)t+1}.
\label{eq:delta}
\end{equation}
In the long time limit $O(t)\sim t^{-1}$, giving $\delta =1$.

Therefore we see that under this simple approximation, the usual mean field active-absorbing transition exponents are obtained.
These are confirmed using numerical simulations as well. A general solution of the dynamics (valid for all $g$ values at all times) is shown
in the Appendix, which in the limiting cases yield the results mentioned above. 

\subsection{Nonuniform guessing of the excess crowd}
In the above mentioned strategy one can find a steady state for any value of the fluctuation. However, unlike
the common practice in MG, the value of $\Delta(t)$ is exactly known to all the agents.
\begin{figure}[tb]
\centering \includegraphics[width=8.5cm]{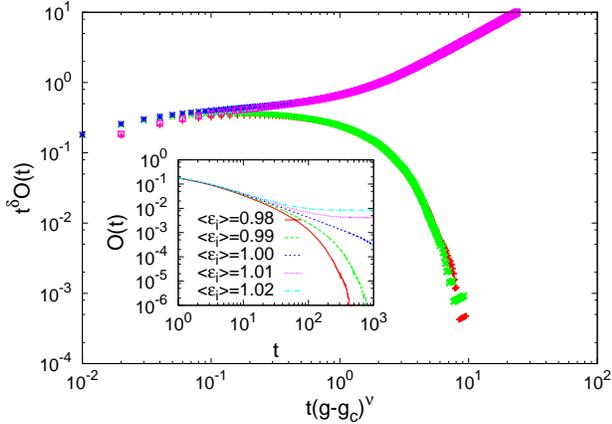}
   \caption{Data collapse for finding $\nu$ in the disordered case for different $x$ values. 
The estimate is $\nu=1.00\pm0.01$. Inset shows the uncollapsed data. The straight line at the critical point gives $\delta=1.00\pm0.01$.
 Simulation data is shown for $M= 10^6$.}
\label{nu-rand}
\end{figure}
\begin{figure}[tb]
\centering \includegraphics[width=8.0cm]{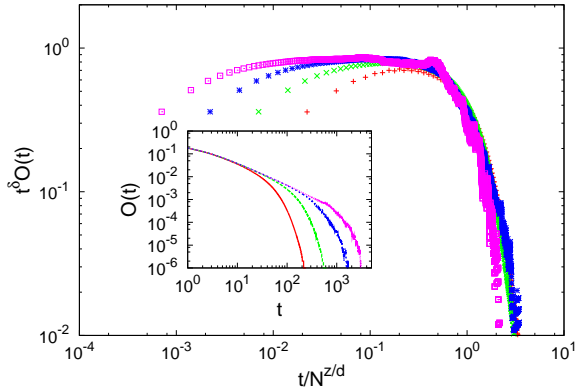}
   \caption{Data collapse for finding $z$ in the disordered case for different system sizes ($M=10^3, 10^4, 10^5, 10^6$) at $x=1.0$. 
The estimate is $z/d=0.50 \pm 0.01$. Inset shows the uncollapsed data. The linear part in the inset confirms $\delta=1.00\pm0.01$.}
\label{z-rand}
\end{figure}
Here we consider the case where each  agent can only make a guess about the value of $\Delta(t)$. Therefore for the $i$-th agent
\begin{equation}
\Delta_i(t)=\Delta (t) (1+\epsilon_i),
\label{eq:dis}
\end{equation}
where $\epsilon_i$ is a uniformly distributed random number in the range $[0:2x]$ and is an annealed variable (i.e., changes at each time step randomly)
with  $|x| \ge 0$. 
Clearly,
\begin{equation}
\langle \Delta_i(t)\rangle=\Delta(t)(1+\langle\epsilon_i\rangle)=\Delta(t)(1+x),
\end{equation}
where $\langle Q\rangle$ ($=1/2x\int\limits_0^{2x}Qd\epsilon$) denotes average of $Q$ over randomness.
With the analogy of the previous case, we expect a transition from zero to finite activity near $x_c=1$.
In Fig.~{\ref{op}} we plot the steady state values of the order parameter against $(x+1)$. Clearly, the active-absorbing transition takes place at $x_c=1$
(note that this is the same point $g_c=2$ where the transition for the pure case took place).

Note that irrespective of whether population inversions occurs, one can generally write
\begin{equation}
\Delta(t+1)=|\Delta(t)-S(t)|,
\end{equation}
with
\begin{equation}
S(t)=\left|\left\langle\frac{\Delta(t)(1+\epsilon)}{M+\Delta(t)(1+\epsilon)}\right\rangle\right|.
\end{equation}
This leads to
\begin{equation}
O(t+1)=O(t)\left|\left\langle\frac{\epsilon}{1+(1+\epsilon) O(t)}\right\rangle\right|.
\label{ann}
\end{equation}

First consider the steady state where $O(t+1)=O(t)=O^*$ (above the critical point). After simplification, the above equation reduces to
\begin{equation}
\frac{(1-O^*)2xO^*}{(1+O^*)}=\ln\left[1+\frac{2xO^*}{1+O^*}\right].
\end{equation}
One can numerically compare the solution of this equation with the  simulations, which agrees well (see Fig.~\ref{op}).
A small $O^*$ expansion of the above equation yields $O^*\sim (x-1)$, giving $\beta=1$.

Also, for small $O(t)$ the dynamical equation would yield (at or above the critical point)
\begin{equation}
\frac{dO(t)}{dt}=(x-1)O(t)-xO^2(t).
\end{equation}
Now, by neglecting the square term (in presence of the linear term above the critical point) one would obtain $\nu=1$ and by keeping the
square term (in absence of the linear term at the critical point) one would obtain $\delta=1$. Thus all the
exponents of the pure case are recovered for annealed disorder.

One can numerically verify the above exponent values using the  following scaling form of the order parameter (writing $x+1=g$)
\begin{equation}
O(t)\approx t^{-\delta}\mathcal{F}\left(t^{1/\nu}(g-g_c), t^{d/z}/N\right)
\end{equation}
where $d$ is the space dimension, which we take to be 4 in this mean field limit. At the critical point, the order parameter follows a power-law relaxation $O(t)\sim t^{-\delta}$ (see inset of Fig.~{\ref{nu-rand}} ) with $\delta=1.00\pm 0.01$. 
\begin{figure}[tb]
\includegraphics[width=8.5cm]{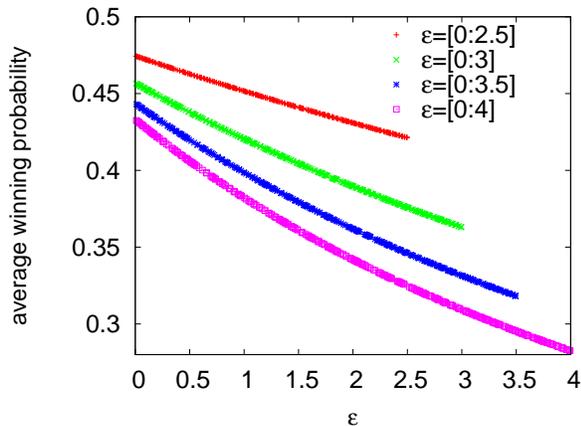}
   \caption{For quenched $\epsilon_i$ (in Eq.~(\ref{eq:dis})) the average pay-offs of the agents are plotted for different $\epsilon$ values having different ranges
as indicated. The monotonic decay with increasing $\epsilon$ clearly indicates that agents with higher 
$\epsilon$ are more likely to be in the majority (see last para of Sec. IIC).}
\label{quench_epsi}
\end{figure}
In Fig.~\ref{nu-rand} we plot $O(t)t^{\delta}$ against $t(g-g_c)^{\nu}$. By knowing $\delta$, $\nu$ can be tuned to get data collapse. The estimate of $\nu$ is $1.00\pm 0.01$. 
Similarly, in Fig.~\ref{z-rand} we plot $O(t)t^{\delta}$ against $t/N^{z/d}$. Again by tuning
$z$, data collapse can be found. The estimate for $z/d$ comes out to be $0.50\pm 0.01$. Thus the analytical estimates are verified and 
the scaling relation $\delta=\beta/\nu$ is satisfied.

We have also tried the case of quenched disorder ($\epsilon$'s are fixed for the agents for all time). Above the critical point, when population
inversion occurs, this would imply that agents with higher $\epsilon$ would change side with higher probability and are more 
likely to be in the majority. A plot of average pay-off for agents having different $\epsilon$ values verify this statement by showing
the monotonic decay of the pay-off with increasing $\epsilon$ (Fig.~{\ref{quench_epsi}}).

\subsection{Following an annealing schedule}
In MG, information about the excess crowd is generally unknown to the agents. In the strategies mentioned above, the excess 
population is known to the agents either exactly or approximately. Here we consider the case, where  knowledge of $\Delta(t)$ is not known to the agents. 

\begin{figure}[tb]
\includegraphics[width=8.5cm]{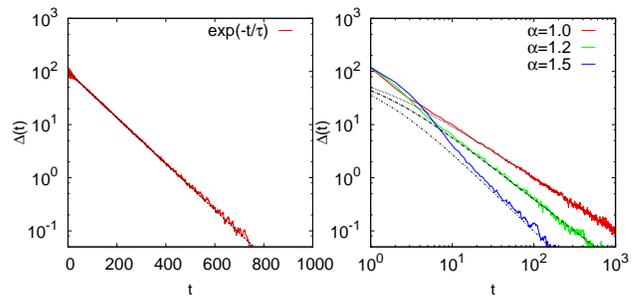}
   \caption{Time variation of the excess population $\Delta(t)$ are plotted for different functional forms of $\Delta^T(t)$. Left: In log-linear scale the 
excess population are plotted for exponential decay. Right: For power law ($\Delta^T(0)/(1+t)^{\alpha}$ decay, with different values of $\alpha$). $M=5\times 10^3$ for the simulations.}
\label{quench}
\end{figure}

The agents follow a simple time evolution function for the time variation of the excess population $\Delta^T(t)$. To begin with, we consider an annealing schedule
\begin{equation}
\Delta^T(t)=\Delta^T(0)\exp(-t/\tau).
\end{equation}
where $\Delta^T(0)$ is taken close to $\sqrt{M}$. In Figs.~{\ref{quench}} we plot the time variation of the actual value of excess 
population as well as $\Delta^T(t)$. We see that $\Delta(t)$ decreases very quickly. Furthermore, it appears that there is a simple relation
between $\Delta^T(t)$ and $\Delta(t)$ such that
\begin{equation}
2\Delta(t)=\Delta^T(t).
\label{eq:rel}
\end{equation} 
This implies that $\Delta(t)\sim 1$ in $\log N$ time. Therefore, in this strategy, even if the actual value of the excess crowd is not supplied 
to the agents, they can find a state where the fluctuation practically vanishes. 
We have also checked whether this is true for some other functions as well. We have taken functional forms such as 
\begin{equation}
\Delta^T(t)=\frac{\Delta^T(0)}{(1+t)^{\alpha}}.
\end{equation}
For all  cases mentioned above we plot (see Figs.~{\ref{quench}} ) $2\Delta(t)$ and $\Delta^T(t)$ to check if they are equal. We conclude that
this relation is not dependent on the functional form of $\Delta^T(t)$ (as long as it is not too fast; see discussions below). 

The response of the order parameter to the assumed trial function can be somewhat  understood as follows: The dynamical equation for $O(t)$ would be
\begin{equation}
O(t+1)=\frac{|\eta(t)-O(t)|}{1+\eta(t)},
\label{qdyn}
\end{equation}
where $\eta(t)=\Delta^T(t)/M$. Considering the case $\eta(t)>O(t)$ (when population inversion takes place) one would arrive at
\begin{equation}
\frac{dO(t)}{dt}-\left(\eta(t)-2\right)O(t)=\eta(t)\left(1-\eta(t)\right).
\end{equation}
A general solution of the above equation will be of the form
\begin{eqnarray}
O(t)&=&\frac{\int\limits_0^tdt_1\eta(t_1)(1-\eta(t_1))e^{\int\limits_0^{t_1}(2-\eta(t_2))dt_2}}{e^{\int\limits_0^t(2-\eta(t_1))dt_1}}\nonumber \\
&&+C_1e^{-\int\limits_0^t(2-\eta(t_1))dt_1},
\label{sol}
\end{eqnarray}
where $C_1$ is a constant. This continuum limit is valid only for the functions $\eta(t)$ that do not decay too fast.
Considering $\eta(t)<2$, one can show that the dominant term of the solution will be of the form
\begin{equation}
O(t)\approx\frac{\eta(t)\left(1-\eta(t)\right)}{2-\eta(t)}\approx\frac{\eta(t)}{2},
\end{equation}
as seen numerically. However, evaluation of Eq.~(\ref{sol}) for $\eta(t)=\eta_0\exp(-t/\tau)$ for $\tau>1/2$ gives 
\begin{equation}
O(t)\sim \frac{\tau}{2\tau-1}\eta(t).
\end{equation}
Therefore, $O(t)\sim \eta(t)/2$ (as in Eq.~(\ref{eq:rel})) is only true when $\tau\gg 1/2$, which is the measure of slowness required in $\eta(t)$ to reduce $O(t)$. 

In the case where $O(0)$ is large or $\eta(t)$ decays too fast (than the limit mentioned above), one would simply have (following Eq.~(\ref{qdyn}))
\begin{equation}
O(t)\sim O(t-1)-\eta (t-1)\sim O(0)-\sum\limits_{k=0}^{t-1}\eta(k).
\label{satu}
\end{equation}
So, $O(t)$ would in fact saturate to a finite value and no population inversion will take place. This is also seen numerically.
\section{Effect of random traders}
According to the strategies mentioned above, if the excess population is known to the agents (which in this case is in fact 
a measure of the stock's price) the fluctuations can have any small value. However, in real markets, there are agents who follow certain strategies depending on the market signal (chartists) and
also some agents may decide completely randomly (random traders). Here we intend to investigate the effect of having random traders in the market, while the rest of
the populations follow the strategies mentioned above.

\subsection*{Single random trader}
When a single random trader is present, even when $\Delta(t)=0$, that trader would choose randomly irrespective of whether he or she is in the minority 
or majority. This will create a crossover between majority and minority with an average time of two time steps. In this way, the asymmetry in the
resource distribution can be avoided. However, that single agent will always be in the majority.

\subsection*{More than one random traders}
\begin{figure}[tb]
\centering \includegraphics[width=9cm]{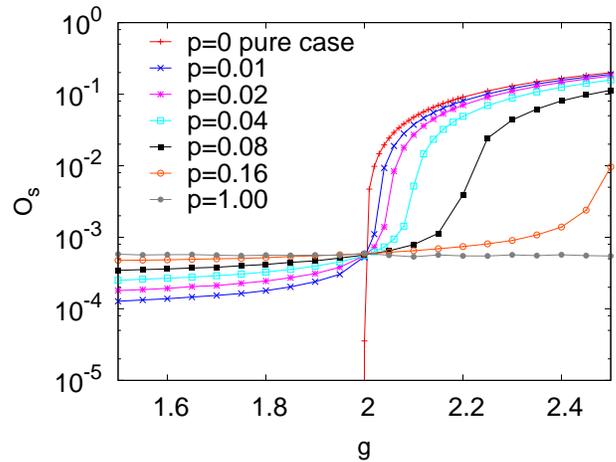}
   \caption{The saturation values of $O_s$ are plotted against $g$ for different fractions $p$ of the random traders. $M=10^6$ for
the simulations.}
\label{noise}
\end{figure}

As is discussed in Sec. II, when all agents follow the strategy described by Eq. (\ref{strategy}), after some initial dynamics, $\Delta(t)=0$ implying that they do not change side at all. However, with a single random trader, in an average time period 2, as he or she selects alternatively between the two choices,  the rest of the population is divided equally between the two choices and it is the random trader who creates the majority by always making himself or herself a loser. This situation can be avoided when there is more than one random trader. In that case, it is not possible always to have them in the majority. There will be configurations where some of the random traders can be in the minority, making their time period of wining to be 2 (due to the symmetry of the two choices). The absorbing state (for $g<g_c$), therefore, never appears with random traders, though the fluctuation becomes non-zero for more than one random traders. However, if the number of random traders ($=pN$, where $p$ is the fraction of random traders) is increased, the fluctuation in the excess population will grow eventually to $N^{1/2}$(see Fig. \ref{noise}). Therefore, the most effective strategy  could be the one in which (i) the fluctuation is minimum and (ii) the average time period of gain will be 2 for all the agents, irrespective of the fact whether they are random traders or chartists. These two are satisfied when the number of random traders is 2.  

Furthermore, if one incorporates the random traders in the strategy described in Sec. IIB, even the knowledge of the excess population will not be exactly needed to reach a state of very small fluctuations.

\section{Summary and Discussions}
In the stochastic strategy minority game, a very efficient strategy is described by Eq.~(\ref{strategy}), where the agents very quickly (in $\log \log N$ time) get divided almost equally ($M$ and $M+1$) between the two choices. This strategy guarantees that a single cheater, who does not follow this strategy, will always be a loser \cite{Dhar:2011}. However, the dynamics in the system stops very quickly, making the resource distribution highly asymmetric (people in the majority stays there for ever) thereby making this strategy socially unacceptable. We present here modifications in the above mentioned strategy to avoid this absorbing state. The presence of a single random trader (who picks between the two choices completely randomly) will avoid this absorbing state and the asymmetric distribution. However, this will always make that random trader a loser. But the presence of more than one random trader will avoid that situation too, making the average time period of switching between majority and minority for all the traders (irrespective of whether they are chatists or random traders) to be 2.  

We show (in Sec.~\ref{sec:2}) that by  varying a parameter, the agents can achieve any value of the
fluctuation. This is an active-absorbing type phase transition and we also find the critical exponents analytically ($\beta=1$, $\nu=1$, $\delta=1$), which are well supported by
numerical simulations.
Then we go on to reduce the knowledge of the agents about the excess populations, which was exactly known to the agents in the earlier strategies. We assume that the agents can only make a guess about the 
excess population. We show using numerical simulations and also using approximate analytical calculations that when  value of the average guess of the agents are not too bad (less than twice the actual value) they can still
reach the state of zero fluctuation. Once again the fluctuation values increase continuously, when the guess becomes worse. This is again an
active-absorbing type phase transition with similar values for critical indices. 
Next we consider the case when the knowledge of excess crowd is completely absent for the agents. In this case the agents assume a time variation (annealing schedule, see Sec. IIC)
for the excess population and they do not look at the actual value. It is shown for several choices of the functional form that the actual value
of the excess population essentially follows the assumed form and thereby goes to zero in finite time (depending on the assumed functional form).
Again we show both analytically and numerically that for slow enough annealing schedule the attainment of zero fluctuation in $\log N$ time can be guaranteed.
Finally, as mentioned before, we also consider the effect of having random traders in the market who decide  absolutely randomly in Sec. III.

We have presented several stochastic strategies for Minority Game. We have shown that 
utilizing the existence of a continuous transition (from $\Delta_s=0$ to $\Delta_s \ne 0$) the agents can find 
the state of zero or arbitrarily small amount of fluctuation (thereby making the system socially efficient) in $\log N$ time with or without knowing the excess population.
Presence or evolution of a few (minimum 2) random traders in the population will not only help the entire population to get out of the absorbing state (for $g<g_c$), but also will have a minimal fluctuation or maximum social efficiency. 

Stated briefly, as a population of $N$ agents try to evolve a strategy such that each of them individually belong to minority among the two choices $A$ and $B$, $N-2$ of them will follow the stochastic strategy given by Eq.~(\ref{strategy}), while the rest 2 will follow a purely random strategy. This will ensure, as shown here, that fluctuation $\Delta_s$ ($\sim |N_A-N_B|$) will have arbitrarily small value (giving maximum social efficiency) achieved in $\log N$ time and the absorbing state will never appear while everyone will have an average period of 2 in the minority/majority. Decrease in the number of random traders will either enforce the absorbing state or indefinite stay in the majority for the random trader, while increase in the number beyond 2 will increase $\Delta_s$, eventually converging to its $\sqrt{N}$ value. As shown, even precise value of $\Delta(t)$ in Eq.~(\ref{strategy}) in unnecessary and appropriate guesses (see Sec. IIB) or appropriate annealing of $\Delta(t)$ (see Sec. IIC) can achieve almost the same results.

\begin{acknowledgements}
\noindent We are extremely thankful for the suggestions made  by an anonymous referee for a generalized version of the  Eq.~(\ref{diff1}) and its solution, as outlined in the Appendix and also for pointing out Eqs. (\ref{ann}) and (\ref{satu}). 
\end{acknowledgements}
\vskip0.5cm
\appendix
\section{General solution of the dynamics}
\noindent In the calculations leading to Eqs.~(\ref{diffO}) and (\ref{diff1}) we have assumed that the two choices alternatively
becomes minority and majority (population inversion happens). That of course is our concern while studying the active phase ($g>g_c$). However, as is clearly
seen from the strategy, this population inversion only happens when $g>1$. So, a more general solution of the dynamics valid for all $g$ can be
done as follows: Consider the auxiliary variable $u(t)=|g-1|^t/O(t)$. Putting this in Eq.~(\ref{diff1}) and neglecting $1/M$ term, one arrives at the
recursion relation
\begin{equation}
u(t+1)=u(t)+g|g-1|^t.
\label{recu}
\end{equation}
Clearly,
\begin{equation}
u(t)=u(0)+\frac{g(1-|g-1|^t)}{1-|g-1|},
\end{equation}
which leads to
\begin{equation}
O(t)=\frac{1-|g-1|}{g}\frac{1}{\left(\frac{1-|g-1|}{gO(0)}+1\right)|g-1|^{-t}-1}.
\end{equation}
For the special case of $g=2$ one gets back Eq.~(\ref{eq:delta}). Also, one can always define the time scale for the above equation as
\begin{equation}
\tau\sim\frac{1}{|\ln(|g-1|)|}.
\end{equation}
Near the critical point $g\to g_c=2$, the already obtained power-law divergence ($(g-g_c)^{-1}$) is recovered. 

Finally, for $g<1$, the dynamical equation will reduce to
\begin{equation}
O(t)\sim \frac{O(0)}{O(0)+1}(1-g)^t.
\end{equation}


\end{document}